\documentclass[draft]{agujournal2019}
\usepackage{url} 
\usepackage{lineno}
\usepackage[inline]{trackchanges}
\usepackage{soul}
\usepackage{graphicx}
\usepackage{amsmath}
\usepackage{makecell}
\usepackage{amssymb}
\usepackage{multirow}
\usepackage{multirow}
\usepackage{color}

\draftfalse

\journalname{JGR: Planets}

\begin{document}
\title{The Heterogeneous Surface of Asteroid (16) Psyche}

\authors{Saverio Cambioni\affil{1}, Katherine de Kleer\affil{2}, Michael Shepard\affil{3}}

\affiliation{1}{Department of Earth, Atmospheric \& Planetary Sciences, Massachusetts Institute of Technology, Cambridge, MA, USA}
\affiliation{2}{Division of Geological \& Planetary Sciences, California Institute of Technology, Pasadena, CA, USA}
\affiliation{3}{Department of Environmental, Geographical \& Geological Sciences, Bloomsburg University, Bloomsburg, PA, USA}

\correspondingauthor{Saverio Cambioni}{cambioni@mit.edu}

\begin{keypoints}
\item Longitude-latitude maps of the thermal inertia and dielectric constant of asteroid (16) Psyche derived from ALMA data.

\item The surface of Psyche is heterogeneous and shows signatures suggestive of both metal- and silicate-rich materials.

\item A depression on Psyche with distinctive thermal-inertia properties suggests an evolved surface processed by impacts.
\end{keypoints}

\begin{abstract}
Main-belt asteroid (16) Psyche is the largest M-type asteroid, a class of object classically thought to be the metal cores of differentiated planetesimals and the parent bodies of the iron meteorites. \citeA{deKleer_2021_Psyche} presented new data from the Atacama Large Millimiter Array (ALMA), from which they derived a global best-fit thermal inertia and dielectric constant for Psyche, proxies for regolith particle size, porosity, and/or metal content, and observed thermal anomalies that could not be explained by surface albedo variations only. Motivated by this, here we fit a model to the same ALMA dataset that allows dielectric constant and thermal inertia to vary across the surface. We find that Psyche has a heterogeneous surface in both dielectric constant and thermal inertia but, intriguingly, we do not observe a direct correlation between these two properties over the surface. We explain the heterogeneity in dielectric constant as being due to variations in the relative abundance of metal and silicates. Furthermore, we observe that the lowlands of a large depression in Psyche's shape have distinctly lower thermal inertia than the surrounding highlands. We propose that the latter could be explained by a thin mantle of fine regolith, fractured bedrock, and/or implanted silicate-rich materials covering an otherwise metal-rich surface. All these scenarios are indicative of a collisionally evolved world.
\end{abstract}

\section*{Plain Language Summary}

Asteroid (16) Psyche is the target of the eponymous NASA mission, which will assess whether the asteroid is the exposed core of an early pre-planet. The Atacama Large (Sub-)Millimeter Array located on the Chajnantor Plateau in Chile allowed the acquisition of temperature images of Psyche at a resolution of 30 km/pixel (the highest ever achieved from Earth). We analyze these images to map the metal content of the first millimeters of the surface as function of longitude and latitude. We find that the relative abundance of metals and silicates varies across the surface. Additionally, we observe that the lowlands of a large depression change temperature much faster than the surrounding highlands as Psyche rotates on its spin axis. Based on this, we propose that the lowlands could exhibit ponds of fine-grained materials, be highly fractured, and/or host silicate-rich materials implanted by impacts. All these scenarios are indicative of an evolved surface processed by impacts.\\ 
~\\
\noindent
\textbf{Keywords}: (16) Psyche; Asteroids -- Surfaces; Remote Sensing -- ALMA; Origin and Evolution

\section{Introduction} \label{sec:intro}

The upcoming NASA Psyche mission will explore Main Belt asteroid (16) Psyche (henceforth just ``Psyche''), an M-type asteroid of about 220 km in size which could have been part of a differentiated planetesimal \cite{2022SSRvElkinsTanton}. Planetesimals are thought to be born 100 km in size and larger \cite<e.g.,>{2009IcarMorbidelli}. During the evolution of the solar system, some may have survived relatively intact, while others suffered collisions and created asteroid families \cite<e.g.,>{2017SciDelbo}, and others may have lost their mantle materials with no apparent associated family \cite{1999IcarDavis}. Psyche could be a differentiated planetesimal that lost its mantle in a hit-and-run collision \cite<e.g.,>{2014NatGeAsphaug}, a rubble pile of material that was previously part of a differentiated body, or a primitive object of unusually metal-rich composition. Answering the question `what is Psyche?' is valuable for constraining the collisional evolution of the planetesimals and requires linking remote sensing observations of their surfaces to the properties of their interiors. 

\begin{sidewaystable}
\begin{center}
\caption{Properties of some M-type asteroids \label{tbl:properties_Mtypes}}
\begin{tabular}{l|rr|rrrrr}
Asteroid &	a [AU] & D [km] & Radar albedo & $\Gamma$ [tiu] & Analog? & 3-$\mu m$? & References\\
\hline
16 Psyche       &   2.92   &   222$^{+4}_{-1}$              &   0.34$\pm$0.08   &       280$\pm$100         &   see text                &   yes     &   S21; DK21; T17\\
21 Lutetia      &   2.44   &   98$\pm$2                     &  	0.24$\pm$0.07   &       20--30              &   EC                      &   yes     &   S15; C11; S10; R11 \\
22 Kalliope	    &   2.91   &   168$\pm$3                    &   0.18$\pm$0.05   & 	    125$\pm$125	        &   IM+PYR                  &   yes     &   S15; M12; F10; OB10 \\
55 Pandora	    &   2.76   &   85$\pm$3                     &   -               &	    -                   &   IM                      &   yes     &   OB10\\	
69 Hesperia	    &   2.98   &.  136$\pm$14                   &   0.45$\pm$0.12   &       -                   &	IM+PYR                  &   yes     &   S15; H05; L15 \\	
77 Frigga	    &   2.67   &   61.4$\pm$0.2                 &   0.14$\pm$0.04   &       -                   &	-                       &   -       &   S15\\			
92 Undina	    &   3.19   &   108$\pm$5                    &   0.38$\pm$0.09   &		-                   &   IM+PYR                  &   yes     &   S15; F11; R00\\	
97 Klotho	    &   2.67   &   88$\pm$24                    &   0.26$\pm$0.05   &	    -                   &   EC, CH                  &   no      &   S15; S10; OB10\\
110 Lydia	    &   2.73   &   88$\pm$3                     &   0.37$\pm$0.10 	&       135$\pm$65          &	IM+PYR                  &   yes     &   S15; DT09; H05; OB10\\
125 Liberatrix  &   2.74   &   48.4$\pm$0.5                 &	-               &		71$^{+26}_{-24}$    &	IM+PYR                  &   no      &   ME21; H05; R00\\
129 Antigone	&   2.87   &   128.7$\pm$0.6                &   0.36$\pm$0.09   &		-                   &   CB?                     &   yes     &   S15; S10; OB10\\
135 Hertha	    &   2.43   &   71$\pm$3                     &   0.18$\pm$0.05 	&		-                   &   CH, EC                  &   yes     &   S15; S10; OB10\\
136 Austria     &   2.29   &   36.9$\pm$0.5                 &   -               &	    -                   &   -                       &   yes     &   L15\\
201 Penelope	&   2.68   &   86$\pm$3                     &   0.40$\pm$0.10   &       $<$50               &   IM+PYR                  &   yes     &   S15; M05; H05; R00\\
216 Kleopatra	&   2.79   &   119$\pm$3                    &   0.43$\pm$0.10   &       $>$50               &   IM+PYR                  &   no      &   S18; M05; OB10; L15\\
224 Oceana	    &   2.65   &   58.2$\pm$0.8                 &   0.25$\pm$0.10 	&		-                   &   EC, CH                  &   -       &   S15; S10\\
250 Bettina     &   3.15   &   121$\pm$2                    &   -               &	    -                   &   IM+PYR                  &   -       &   OB10; H07 \\	
347 Pariana	    &   2.61   &   48.6$\pm$0.1                 &   0.36$\pm$0.09 	&		-                   &   IM, CB                  &   -       &   S15; S10 \\
382 Dodona      &   3.12   &   65.2$\pm$0.5                 &   -               &		60$^{+90}_{-45}$  	&   IM+PYR                  &   -       &   ME21; H11\\
413 Edburga	    &   2.58   &   25$\pm$5                     &   0.35$\pm$0.09 	&       100$^{+60}_{-60}$   &   IM, EC                  &   no      &   S15; ME21; H07; R00 \\
441 Bathilde    &   2.81   &   66$\pm$2                     &	0.20$\pm$0.05  	&      	180$^{+20}_{-60}$   &   -                       &   -       &   S15; PG19\\
497 Iva	        &   2.85   &   40.9$\pm$0.3                 &   0.24$\pm$0.08 	&	    70$^{+19}_{-25}$	&   SIM, CH                 &   no      &   S15; ME21; OB10; OB10\\
766 Moguntia    &   3.02   &   41.0$\pm$0.1                 &	-               &		-                   &	IM+OLI                  &   -       &   H11\\
785 Zwetana	    &   2.57   &   49$\pm$1                     &   0.26$\pm$0.07	&	    $<$50               &   IM, CB                  &   no      &   S15; M05; OB10; OB10\\
798 Ruth        &   3.01   &   46$\pm$7                     &	-               &		-                   &	IM+OLI                  &   -       &   H11 \\
1210 Morosovia  &   3.01   &   33.7$\pm$0.3                 &	-               &  		-                   &	IM+OLI                  &   -       &   H11\\
\end{tabular}
\end{center}
\caption{``a'' is the semi-major axis of the asteroid in astronomical units (AU). D is the mean or effective diameter, latest WISE value from \citeA{2011ApJMasiero,2012ApJMasiero,2014ApJMasiero,2021PSJMasiero} as listed in the Minor Planet Physical Properties Catalogue (\url{https://mp3c.oca.eu/}), except for Psyche, (21) Lutetia, and (69) Hesperia and (798) Ruth, for which the values are from \citeA{shepard2021}, \citeA{2011SciSierks}, and \citeA{2018AALagoa}, respectively. $\Gamma$ is thermal inertia, that is, the resistance of surface material to change temperature. Uncertainties are 1-standard deviation. IM = iron meteorites; EC = enstatite chondrite; CB = high-iron carbonaceous chondrites; CH = CH meteorites; SIM = stony-iron meteorites. The acronyms PYR and OLI indicate the detection of pyroxene- and olivine-bands in the bodies' spectra. References for radar albedo, thermal inertia,  silicates and OH/H$_2$O band are listed using the following acronyms:  C11: \citeA{2011SciCoradini}; DK21: \citeA{deKleer_2021_Psyche}; DT09: \citeA{2009PSSDelbo}; F10: \citeA{2010IcarFornasier}; F11: \citeA{2011IcarFornasier}; H05: \citeA{2005IcarHardersen}; H07: \citeA{2007LPIHardersen}; H11: \citeA{2011MHardersen};  L15: \citeA{2015IcarLandsman}; M05: \citeA{2005DPSMueller}; M12: \citeA{marchis2012multiple}; ME21: \citeA{2021PSJMacLennan}; OB10: \citeA{2010IcarOckert}; PG19: \citeA{2020APodlewska} R11: \citeA{2011IcarRivkin}; S10: \citeA{shepard2010}; S15: \citeA{shepard2015}; S18: \citeA{2018IcarShepard}; S21: \citeA{shepard2021}; T17: \citeA{takir2017}.}
\end{sidewaystable}

The M-type asteroids \cite{tholen1984} are a diverse group (Table \ref{tbl:properties_Mtypes}), which is broken down in the Bus--DeMeo taxonomy into the Xc class, associated with enstatite chondrites, and the Xk class, associated with mesosiderites \cite{demeo2009}. \citeA{2014IcarNeeley} compared the optical-near-infrared spectra for 29 M-type asteroids to meteorites and found that, while the best analog types are iron meteorites in most cases, 21\% of the targets were better fit by enstatite chondrites, consistent with early studies \cite<e.g.,>{chapman1973,2009IcaVernazza}. Insights into the concentration of metal in the near-surface of asteroids come from measurement of radar echoes, in which a circularly polarized continuous wave signal is trasmitted to the asteroid and the echo power in the same (SC) and opposite (OC) senses of polarization is measured. The OC radar albedo (henceforth just ``radar albedo'') is the ratio of the power received from target asteroids relative to that which would be measured from a metallic sphere of the same cross-sectional area and at the same distance. Radar albedo increases with increasing metal content of the near-surface material \cite<e.g.,>{shepard2015}. The average radar albedo of the M-type group is 0.28 $\pm$ 0.13 \cite{shepard2010,shepard2021}, which is much higher than that of S-type or C-type asteroids \cite<$\sim$ 0.14, e.g.,>{2007IcarMagri} and, as such, indicative of a higher metallic content of the surface material \cite{ostro1985}. Another proxy for surface metal content is the thermal inertia $\Gamma$, which is the resistance of the surface material to change temperature during the diurnal illumination cycle. $\Gamma$ is a composite parameter function of the material thermal conductivity $\kappa$, specific heat capacity $c_p$ and bulk density $\rho$ of the surface ($\Gamma = \sqrt{\kappa c_p \rho}$). Thermal inertia $\Gamma$ is measured in units of J m$^{-2}$ K$^{-1}$ s$^{-0.5}$, which will hereafter be referred to as ``tiu'' for ``thermal inertia unit''. For a given particle size, metal-rich regolith tend to have higher thermal inertia than silicate-rich ones \cite{matter2013}; consistently, some M-type asteroids appear to have higher thermal inertia than similar-sized C-type or S-type asteroids \cite<e.g.,>{delbo2015}. However, evidence for a correlation between thermal inertia and radar albedo among M-type asteroids is marginal to none \cite<Table \ref{tbl:properties_Mtypes}; see also>{landsmanevidence,elkins-tanton2020}.

Spectroscopic observations of M-type asteroids also revealed the presence of silicate features in otherwise featureless iron-meteorite-like spectra \cite<e.g.,>{2010IcarFornasier,hardersen2005}. This material may be what remains of stripped rocky mantles, supporting the hypothesis that M-type asteroids are the cores of differentiated planetesimals \cite<e.g.,>{bell1989,2014NatGeAsphaug,2015asteScott}. At the low pressures of planetesimals' magma oceans, olivine is expected to be the first mineral to crystallize across a vast range of possible bulk compositions \cite{2013LPIElkins, elkins2017planetesimals}. Although the spectroscopic signature of olivine was observed on M-type asteroids (766) Moguntia, (798) Ruth, and (1210) Morosovia \cite{2011MHardersen}, the near-infrared spectra of many other M-type asteroids, such as Psyche, (22) Kalliope, (69) Hesperia, (110) Lydia, and (201) Penelope were found to show only a weak absorption feature near 0.93 $\mu m$ indicative of low-Fe, low-Ca pyroxene \cite{2004AJClark,2005IcarHardersen,2010IcarFornasier,2007AABirlan,sanchez2017}. Pyroxenes may form through reduction of olivine in a low oxygen fugacity environment, such as that of the nebular region close to the Sun or in a forming planetary core \cite{2005IcarHardersen,2013LPIElkins}. The dominance of one mineralogy over the other in the M-type population could be indicative of where the asteroids accreted or how they differentiated \cite{2005IcarHardersen}.

The third component detected on the surfaces of M-type asteroids is hydrated minerals \cite<e.g.,>{rivkin2000nature,shepard2015,takir2017}. If hydrated minerals are endogenic to M-types, this would suggest that they accreted beyond the water-ice condensation line of the early solar system \cite{2015IcarLandsman,2022SSRvElkinsTanton}. Another explanation is that the hydrated material is exogenic \cite{2002SoSyRBusarev,shepard2015,avdellidou2018}, consistent with predictions of low volatile content in silicate magmas on planetesimals \cite{weiss2013differentiated} and with spacecraft observations of implanted exogenic material on asteroids of sizes ranging from 500-km-sized Vesta to sub-km-sized (101955) Bennu \cite{2012IcarReddy,2021NatAsDellaGiustina}. Another possible explanation is that space weathering of M-type asteroids' surface material may produce hydrated minerals via interaction of solar wind protons with oxygen-bearing minerals \cite{2015IcarLandsman}, as was proposed for the Moon \cite<e.g.,>{2009SciClark,2009SciSunshine,2009SciPieters}; although, more laboratory experiments are needed to test this proposal specifically for M-types.

\subsection{This work}

On Psyche, variations in metal and silicate content have been proposed on the basis of rotational variations of the surface at visible, infrared and radar wavelengths \cite{viikinkoski2018,ferrais2020,hardersen2005,sanchez2017,takir2017,shepard2021}. \citeA{deKleer_2021_Psyche} observed Psyche in thermal emission using the Atacama Large (Sub-)Millimeter Array (ALMA) at a resolution of 30 km over 2/3 of its rotation and fit for the thermal inertia and dielectric constant (that is, the polarizability of the surface material, to be presented in Section \ref{sec:intro_param}) with a single value of each property over the entire surface. Their results showed the presence of thermal anomalies which support the heterogeneity of surface composition and could not be explained by surface albedo variations only. Motivated by this, here we fit a model to the same ALMA dataset that allows thermal inertia and dielectric constant to vary across
the surface. We describe the methodology in Section \ref{sec:method}. In Section \ref{sec:results}, we present the results as longitude-latitude maps of surface properties and test their robustness to model assumptions. In Section \ref{sec:discussion}, we discuss possible metal-rich and metal-poor areas on the basis of dielectric constant measurements and discuss a prominent feature in the thermal inertia map.
\section{Materials and Methods} \label{sec:method}

\subsection{Data} \label{sec:data}

The data analyzed in this paper were obtained at ALMA on 2019 June 19 between 06:33 and 09:11 UT.  The observations, data reduction, and imaging methods are described in \citeA{deKleer_2021_Psyche} and are only briefly summarized here. Data were obtained at a wavelength of 1.3 mm over 8 GHz of total bandwidth, all of which was used in the analysis. The maximum baseline at the time of observation was 16.2 km, and Psyche was at a distance of 2.04 AU from Earth (and 2.78 AU from the Sun) resulting in a spatial resolution of $\sim$30 km on Psyche, or an angular resolution of $\sim$0.02''. The sub-observer and sub-solar latitudes on Psyche at the time of observation were -14$^{\circ}$ and 3$^{\circ}$ respectively, and the solar phase angle was 17$^{\circ}$. Rotation curves of the total flux density and peak brightness temperature for the ALMA observations are in Figure 2 of \citeA{deKleer_2021_Psyche}.
The data were reduced and calibrated by the ALMA pipeline, and were further calibrated and imaged using the Common Astronomy Software Applications (CASA) package \cite{mcmullin2007}. The duration of each ALMA scan of Psyche was under 1 minute, and several scans were imaged together to improve the signal-to-noise. Loss of resolution from smearing was avoided by keeping the observing interval of jointly-imaged scans below 6 minutes. Each final image has a signal-to-noise of $\sim$45-50 and a noise level around 2.5 K. 

\subsection{Choice of model parameters}
\label{sec:intro_param}

Measurements of thermal emission of asteroids provide constraints on the particle size, porosity, and/or metal content of the surface materials. Thermal observations from Earth have been typically acquired in infrared wavelengths and interpreted using thermophysical models \cite<e.g.,>{delbo2015} for which the free parameters are the thermal inertia $\Gamma$ (defined in Section \ref{sec:intro}) and the roughness of the surface $f$. The latter measures the root-mean-square slope of topography that has spatial scales below the resolution of the shape model of the asteroid. A rougher surface has higher mean surface slope and stays warmer for longer periods of time than it would if it were smoother due to higher absorption of sunlight and more self-heating \cite<e.g.,>{delbo2015,2017MNRASRozitis}. Different approaches to model surface roughness exist in literature \cite{delbo2015}. The thermophysical model we use here (to be presented in Section \ref{sec:method_images}) allows for modelling of surface roughness by carving one spherical craters onto each facet of the shape model and tune the root-mean-square slope $f$ of the facets by either changing the opening angle or the areal fractional coverage of the craters, or both.

Mid-infrared observations do not currently resolve the thermal emission from the surface of Psyche because the high required angular resolution ($\sim$ 0.03'') is beyond the capabilities of existing facilities. Alternatively, \citeA{deKleer_2021_Psyche} used ALMA to acquire thermal emission maps of Psyche at a resolution of 30km/pixel. ALMA is an astronomical interferometer composed of 66 antennas distributed in the Chajnantor plateau in Chile. The antennas observed Psyche's thermal emission at millimiter wavelengths and combined the signals interferometrically into a synthetic signal equivalent to that observed by a putative telescope with an effective diameter of 16 kilometers. The foundational theory of how to analyze thermal emission of solar system bodies at millimeter wavelengths can be found in numerous investigations of the Moon \cite<e.g.,>{1949AuSRAPiddington,1972PrAAMuhleman,1974MoonUlich,2019IcarZheng}, dwarf planet Ceres \cite{1988AJWebster,1989PASPWebster,redman1998,1994AAAltenhoff,2020AJLi}, asteroids (3) Juno \cite{1994AAAltenhoff,junoalma}, (4) Vesta \cite<e.g.,>{1989PASPWebster,1994AAAltenhoff}, Psyche \cite{deKleer_2021_Psyche}, (21) Lutetia \cite{gulkis2012} and (2867) Steins \cite{gulkis2010}, and of icy solar system bodies \cite{muhleman1991,lellouch2017,trumbo2018,dekleer2021}. Discussion of issues related to physical propagation of thermal emission at different wavelengths can be found in \citeA{lagerros1996,redman1998,keihm1982,keihm2013,2017JGREHayne}, and in \citeA{2021BAASKleer} for the specific case of solar system science with ALMA.

At millimeter wavelengths, the thermal emission of the surface (or, equivalently, its brightness temperature) is sensitive to the complex dielectric constant $\epsilon$ of the near surface material and the temperature profile within the (sub)surface. The dielectric constant measures the polarizability of the surface material, has real part $\epsilon'$ and imaginary part $\epsilon''$ (as such, $\epsilon = \epsilon' + i \epsilon''$) and is equal to $\tilde{n}^2$, where $\tilde{n} = n + i K$ is the complex refractive index. $n$ is the ratio of the speed of light in a vacuum to the speed of light in the material and $K$ is the extinction coefficient measuring the amount of attenuation when the electromagnetic wave propagates through the material. The extinction coefficient defines the electrical skin depth $\delta_{elec} = \lambda/(4 \pi K)$ within which the emission is reduced by 1/e assuming a non-conducting material for emission at wavelength $\lambda$. While this depth is small at infrared wavelengths, it becomes significant at longer wavelengths, hence the need to integrate the thermal emission within the subsurface \cite<e.g.,>{gulkis2010} for a temperature profile as a function of thermal inertia $\Gamma$ and surface roughness $f$. Since $\tilde{n}$ defines the reflection coefficients of the surface, the millimeter emissivity of the surface $E$, that is, the degree of emission relative to a blackbody, is a decreasing function of dielectric constant. For low-emissivity asteroids like Psyche \cite<global $E$ = 0.4 -- 0.7,>{deKleer_2021_Psyche}, thermal emission at millimiter wavelengths is particularly sensitive to metal content and porosity of the surface material (see their section 3.5.3).

\subsection{Thermal emission images} \label{sec:method_images}

To interpret the ALMA observations, we generate thermal emission images at the same viewing geometry and wavelength of the ALMA data with thermal inertia and dielectric constant as free parameters. The thermophysical simulation setup and the methodology to create model images are the same of \citeA{deKleer_2021_Psyche}, with a few adaptions as described in the following.

We use a well-established ThermoPhysical Model \cite<TPM,>{delbo2015} to solve the non-dimensionalized 1-D heat conduction equation with thermal inertia $\Gamma$ as free parameter. We explore the following values of thermal inertia: $\Gamma =$ (25,   75,  116,  135,  156,  181,  210,  283,  442,  594, 1000) tiu. We assume surface roughness $f$ = 0$^\circ$ (that is, a smooth surface) as this was found to best-fit the thermal emission data at both infrared \cite{matter2013,landsman2018} and mm-wavelengths \cite{deKleer_2021_Psyche}. We assume bolometric emissivity $E_{bol} = 0.9$, bond albedo $A$ = 0.05, and asteroid topography as defined by the shape model by \citeA<>{shepard2021}. We test the robustness of the results against the assumption that the facets of the shape model are smooth and  the entire surface has the same Bond albedo in Section \ref{sec:robustness}.

Next, we compute the blackbody flux density for the facets of the shape model based on the temperature output of the TPM. For each facet of the shape model, the TPM computes the temperature as a function of dimensionless depth parameter x' = x/$\delta_{th}$ , where x is the depth in physical units and $\delta_{th}$ is the diurnal thermal skin depth over which the amplitude of the diurnal temperature wave is reduced by 1/e:

\begin{equation}
\label{eq:d_thermal}
    \delta_{th} = \sqrt{\frac{P}{\pi}}\frac{\Gamma}{\rho c_p}
\end{equation}
\noindent
with rotation period $P$ = 4.195948 hours for Psyche \cite{shepard2021}. For each value of $\Gamma$, we compute the corresponding $\delta_{th}$ assuming a bulk density $\rho = $ 3500 kg/m$^3$ which was derived from Psyche's radar albedo of 0.34 $\pm$ 0.08 by \citeA{shepard2021} and a value of specific heat $c_p = $ 370 J/kg K. The latter is an average value of possible meteorite analogs for Psyche identified by \citeA{elkins-tanton2020}: mesosiderites ($c_p = $ 383 $\pm$ 6 J/kg K), iron IIAB ($c_p = $ 342 $\pm$ 6 J/kg K) and iron IIIAB ($c_p = $ 375 $\pm$ 23 J/kg K) computed at $\sim$200 K \cite{2013Consolmagno,2013ConsolmagnoSchaefer}. We convert the dimensionless depth parameter x' to physical units and calculate the temperature as a function of physical depth. We use Eq. 3 from \citeA{dekleer2021} to integrate the blackbody flux density through the subsurface along the viewing path. The integration along the viewing path depends on the dielectric constant $\epsilon$. Because $K^2/\epsilon'<$ 0.006 for most meteorites \cite{campbell1969}, hereafter we make the simplification that $\epsilon = \epsilon'$ as in \citeA{deKleer_2021_Psyche}. We explore the following values of $\epsilon$: (3.0, 7.5, 12.0, 16.5, 17.5, 18.5, 19.5, 20.5, 21.5, 25.5, 30.0, 55.0, 80.0). This array is more finely sampled around the best-fit $\epsilon$ = 18.5 found by \citeA{deKleer_2021_Psyche}. We discuss the range of applicability of the results based on the assumed bulk density value in Section \ref{sub:metal_silicate}.

Next, we correct the thermal emission from each facet for the millimeter emissivity $E$ of the surface material. The emissivity as a function of emission angle is computed using the Fresnel reflection coefficients for polarization parallel and perpendicular to the direction of propagation, which in turn depend on the complex refractive index \cite<see Eq. 9 in>{deKleer_2021_Psyche} and thus on the dielectric constant of the material (Section \ref{sec:intro_param}). We explore the same array of values of dielectric constant as described above.

\begin{figure}
\centering
\includegraphics[width=\linewidth]{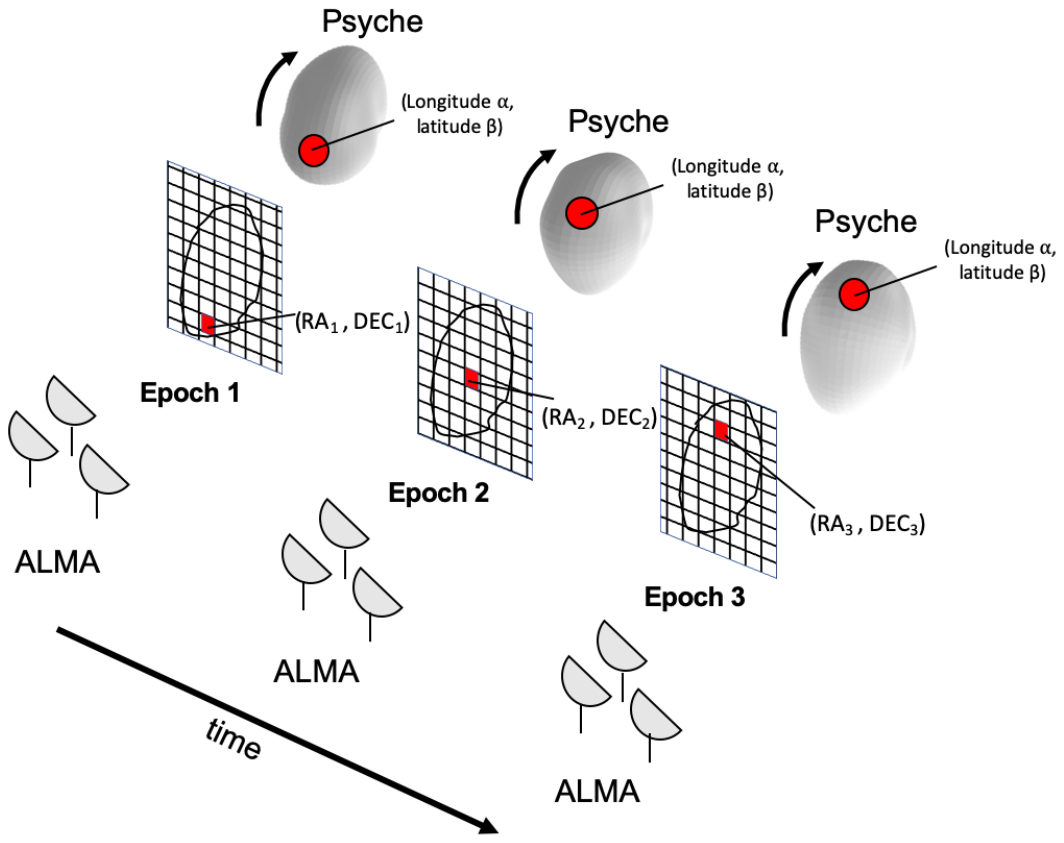}
\caption{As the asteroid rotates during the observation epochs, the same facet of longitude $\alpha$ and latitude $\beta$ is seen at different right ascensions R.A. and declinations decl. in the ALMA images. We use the knowledge of the (R.A., decl.) of the projected facets and the (R.A., decl.) of the ALMA pixels to associate a thermal-emission curve to each facet. We then fit modelled thermal emission curves corresponding to different values of thermal inertia and dielectric constant to the observed thermal emission curves.}
\label{fig:method}
\end{figure}

Next, we obtain model thermal-emission images for Psyche by projecting the shape model by \citeA{shepard2021} onto the plane-of-sky in the International Celestial Reference System (ICRS), which is that of the ALMA observations. We illustrate the projection of the surface areas onto the R.A. and decl. observed by ALMA in Figure \ref{fig:method}. The vertices of the facets are arrays defined in the asteroid reference system, which has positive $\hat{x}$ axis along the direction of the prime meridian as defined by \citeA{shepard2021}, positive $\hat{z}$ axis in the direction of the rotation pole, and the third axis to complete the orthonormal frame. We rotate the vertices to ICRS as described on the DAMIT website (\url{https://astro.troja.mff.cuni.cz/projects/damit/pages/documentation}):

\begin{equation}
    r_{ICRS}=R_x(\theta_3)~ R_z(\theta_1)~R_y(90-\theta_2)~R_z(\phi_0 + \frac{2\pi}{P} (t-t_0))~r_{ast}
\end{equation}
\noindent
where $\theta_1$ = 36$^\circ$ and $\theta_2$ = -8$^\circ$ are the longitude and latitude of the asteroid spin axis, $\phi_0$ = 341.56$^\circ$ is the initial rotation angle of the asteroid, $t$ is the observation epoch, $t_0$ is J2000, and $\theta_3$ = 23.44 $^\circ$ is the Earth's obliquity. The matrices $R_x$, $R_y$ and $R_z$ are defined, for a generic angle $\theta$, as:

\begin{equation}
    R_x(\theta) = 
\begin{bmatrix}
1 & 0 & 0 \\
0 & \cos{\theta} & -\sin{\theta} \\
0 & \sin{\theta}  & \cos{\theta} \\
\end{bmatrix}, R_y(\theta) = 
\begin{bmatrix}
\cos{\theta} & 0 & \sin{\theta}\\
0 & 1 & 0\\
-\sin{\theta}  & 0 & \cos{\theta}\\
\end{bmatrix}, R_z(\theta) = 
\begin{bmatrix}
\cos{\theta} & -\sin{\theta} & 0\\
\sin{\theta}  & \cos{\theta} & 0\\
0 & 0 & 1\\
\end{bmatrix}
\end{equation}

Next, we interpolate the thermal emission values of the projected facets that are visible to ALMA in R.A. and decl. to produce a 2D array whose pixels correspond to those of the ALMA images. Finally, the model images are convolved with the ALMA beam to model the observation conditions. The above procedure is used to build a look-up table of images for every combination of the values of thermal inertia $\Gamma$ and dielectric constant $\epsilon$ listed above.  

\subsection{Fitting $\Gamma$ and $\epsilon$ to the thermal-emission curves}
\label{sec:fit_pixel_space}

Our goal is to fit a value of thermal inertia and a value of dielectric constant to the ALMA images for each surface area, and see whether this can explain the residuals from the best-fit global model by \citeA{deKleer_2021_Psyche}, who determined a global best-fit value of thermal inertia and dielectric constant for the entire surface to the data. Here the term ``surface area'' indicates the ensemble of projected facets of the shape model that are imaged in different ALMA pixels as the asteroid rotates and the surface area moves across the plane of sky. We apply the procedure of Section \ref{sec:method_images} to the data to determine the R.A. and decl. of the projected facets that are visible at a given observing epoch and associate thermal-emission values to such facets based on the corresponding ALMA pixels. The thermal emission curve $\textbf{D}$ of surface area is the variation in thermal emission as a function of time as the asteroid rotates and a facet moves across the plane of sky.  We repeat the same procedure to generate model thermal-emission curves $\textbf{M}(\Gamma,~\epsilon)$ for each facet as a function of thermal inertia $\Gamma$ and dielectric constant $\epsilon$. We evaluate the goodness-of-fit of each modelled thermal emission curve $\textbf{M}(\Gamma,~\epsilon)$ for a given thermal inertia $\Gamma$ and dielectric constant $\epsilon$ by evaluating its $\chi^2_{r}$ (reduced chi-squared) function \cite<e.g.,>{hanuvs2015thermophysical}:

\begin{equation}
\label{eq:chi2}
    \chi^2_{r}(\Gamma,~\epsilon) = \frac{1}{N_{obs}-2} \sum_{i = 1}^{N_{obs}} \frac{(\textbf{M}^i (\Gamma,~\epsilon) - \textbf{D}^i)^2}{\sigma_{\textbf{D}}^2},
\end{equation}
\noindent
where $\sigma_{\textbf{D}}$ is the image background noise (Section \ref{sec:data}) and $N_{obs}\leq22$ is the number of times the projected facet was observed by ALMA. The set $\textbf{S}$ of values of thermal inertia and dielectric constant that best fit the thermal emission curve $\textbf{D}$ are found by taking the median of the $(\Gamma,~\epsilon)$ whose $\chi^2_{r}(\Gamma,~\epsilon)$ values meet the following criterion:

\begin{equation}
\label{eq:chi_stat}
    (\Gamma,~\epsilon)\in \textbf{S}~~if~~\chi^2_r(\Gamma,~\epsilon) < min_{\chi^2_r} \times \bigg(1+\sqrt{\frac{2}{N_{obs}-2}}~\bigg).
\end{equation}

We consider the median to be a more reliable estimator of the central tendency of $\textbf{S}$ than the minimum of the $\chi_r^2$, which is not distinguishable from other solutions in $\textbf{S}$ in virtue of Equation \ref{eq:chi_stat}, the mean of $\textbf{S}$, which is more sensitive to outliers that the median, and the mode of $\textbf{S}$, which is not sensitive to elements in the set different from the most recurrent ones. We compute the uncertainty of the best-fit values of $(\Gamma,~\epsilon)$ as the standard deviation of the solutions in $\textbf{S}$. 

\subsection{Mapping $(\Gamma,~\epsilon)$ in longitude and latitude} \label{sec:lat-lon-maps}

We display the best-fit $(\Gamma,~\epsilon)$ and uncertainties as function of asteroid-centric latitude $\alpha$ and latitude $\beta$ using the Mollweide (equal-area, pseudocylindrical) cartographic map projection technique with the prime meridian (PM) at the center of the map. To take into account the correlation between nearby pixels that characterizes the ALMA images \cite<that is, each ALMA beam contains about 45 pixels,>{deKleer_2021_Psyche}, we bin the surface properties in longitude and latitude on a grid with elements of angular size 5$^\circ \times$5$^\circ$ and apply Gouraud shading \cite{gouraud1971continuous}. We hatch the region between longitudes 60$^\circ$W and 120$^\circ$W to indicate that the solution for that region could be affected by model artifacts. The region was observed at high emission angles in less than 1/3 of the ALMA observations and includes an equatorial area (longitude $\sim$ 90$^\circ$W, latitudes $\pm$ 30$^\circ$) where unmapped topography could be present \cite{shepard2021}. We remove from the maps those areas that have a bad goodness of fit, that is, $\chi^2_r>$10 (the optimal value is $\chi^2_r\sim$1), or $N_{obs}\leq$ 3. The latter condition follows from Equations \ref{eq:chi2}--\ref{eq:chi_stat}, where a surface area shall be visible a number of times at most equal to the number of degrees of freedom plus 1 in order for the inverse problem to be well-conditioned. At least two data points are needed to derive the best-fit value of $\Gamma$ (which controls the variation of temperature as a function of time), and another data point is needed to break the degeneracy on the best-fit value of $\epsilon$. However, if a surface area is observed just for a few epochs near the terminator and then it rotates away from sight, its surface properties may remain unconstrained even if $N_{obs}\geq$ 3. To assess the minimum required number of $N_{obs}$ to fit the data, we test the fitting procedure in Section \ref{sec:fit_pixel_space} to retrieve the surface properties of synthetic Psyche's images with added Gaussian noise based on $\sigma_D$ from the data. We confirm that constraining a solution for $N_{obs}\geq$ 3 is possible everywhere on Psyche despite for a few areas at the south pole, which we remove from the map.

\section{Results} \label{sec:results}

\subsection{Comparison between the global solution and the local solution}

\begin{figure}
\centering
\includegraphics[width=\linewidth]{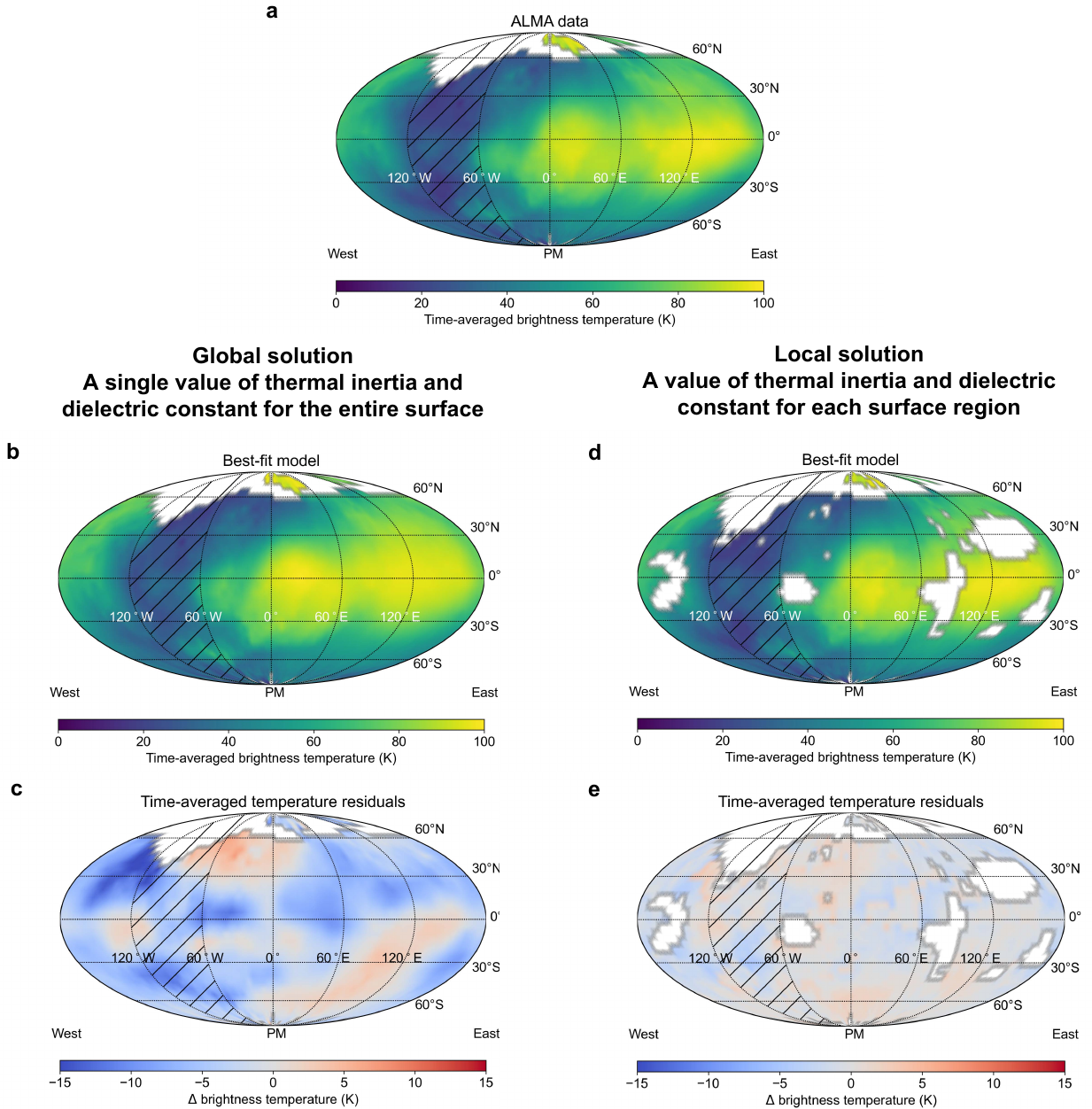}
\caption{\textbf{ALMA data (top panel), global solution (left panels) and local solution (right panels) of the thermal emission of Psyche} a, Time-averaged brightness temperature corresponding to ALMA data. b, Time-averaged brightness temperature corresponding to the best-fit surface properties in \citeA{deKleer_2021_Psyche}. c, Time-averaged residual brightness temperature (that is, time-average of the model minus the data) for the model in panel b. d, Time-averaged brightness temperature corresponding to the best-fit properties for each surface area (this study). e, Time-averaged residual brightness temperature for the model in panel d. In panels d and e, we removed the areas that were either observed less than 3 times or have bad goodness of fit ($\chi^2_r$ $>$ 10). The hatched region is where possible model artifacts may affect the quality of the solution (see Section \ref{sec:lat-lon-maps}). The meridians are spaced 60$^\circ$ apart, so that ``West'' indicates PM$-$180$^\circ$ and ``East'' indicates PM$+$180$^\circ$. The parallels are spaced 30$^\circ$ apart. The sub-observer point is at latitude $-14^\circ$, around which the resolution of the maps is estimated to be $\sim 5^\circ$.}
\label{fig:data_model}
\end{figure}

Figure \ref{fig:data_model}a is the map of brightness temperature corresponding to the time-averaged thermal emission observed with ALMA, that is, the brightness temperature corresponding to the observed thermal emission averaged across the different epochs when the location was visible. Figure \ref{fig:data_model}b and \ref{fig:data_model}c are maps of the time-averaged model brightness temperature and time-averaged residual brightness temperature, respectively, for a smooth surface with $\Gamma=$ 283 tiu and $\epsilon=$ 18.5 which is the global best-fit case from \citeA{deKleer_2021_Psyche}. Figure \ref{fig:data_model}d and \ref{fig:data_model}e are the maps of the time-averaged model brightness temperature and time-averaged residual brightness temperature obtained by fitting a value of $\Gamma$ and $\epsilon$ to the thermal emission curve of each area on Psyche (this study). Hereafter, we will refer to the results in panels b and c as the ``global solution" and those in panels d and e as the ``local solution''. We find that the structures in the temperature residual map of the global solution (Figure \ref{fig:data_model}c) 
are well explained by a variegation in thermal inertia and dielectric constant over the surface: the residuals for the local solution (Figure \ref{fig:data_model}e) are randomly distributed over the surface with mean value and standard deviation equal to -0.3 K and 1.3 K, respectively.

\subsection{Maps of thermal inertia and dielectric constant}
\label{ref:surface_properties}

%
\begin{figure}
\centering
\includegraphics[width=\linewidth]{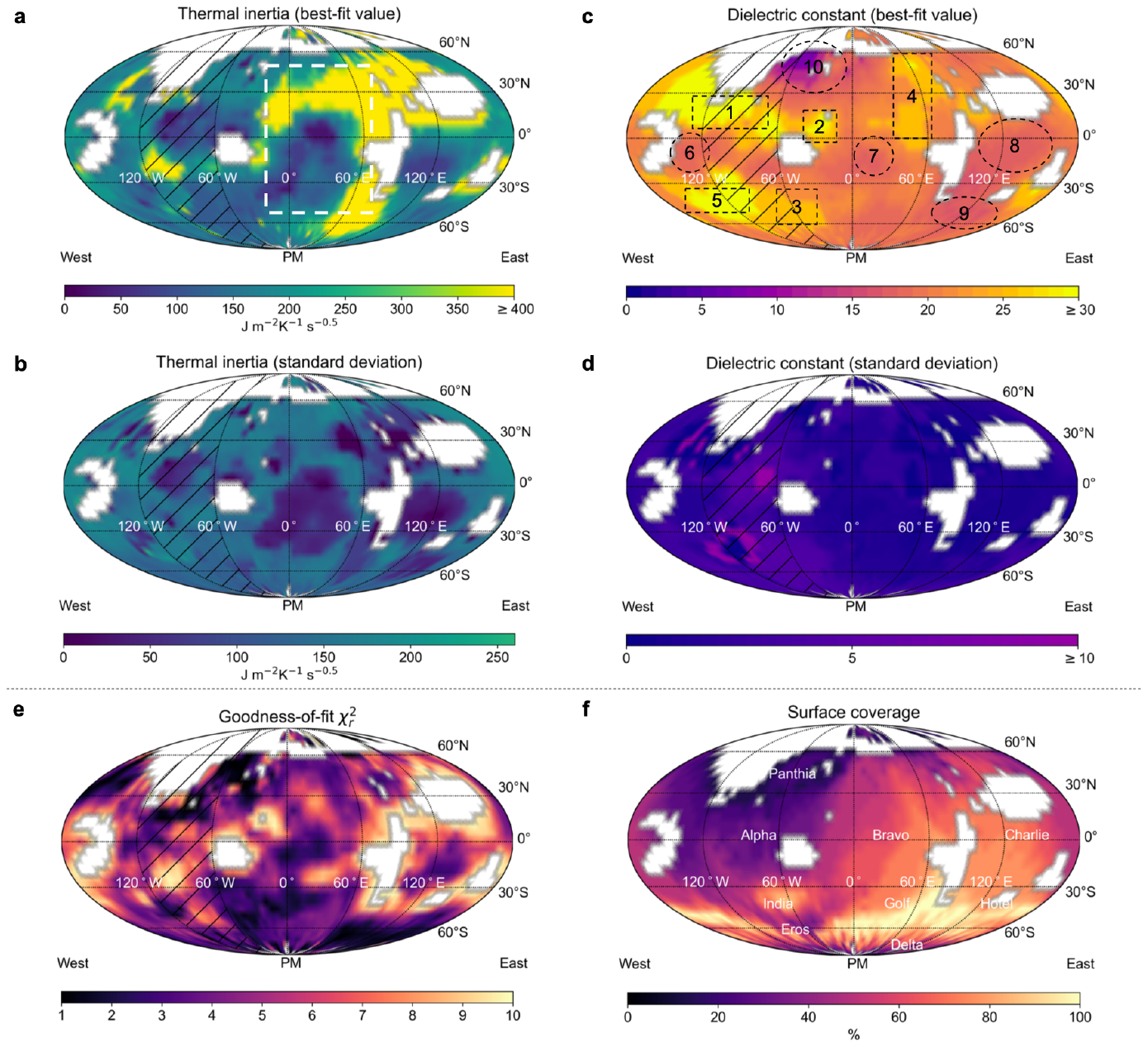}
\caption{\textbf{Maps of thermal inertia and dielectric constant of asteroid (16) Psyche derived from the ALMA thermal emission data.} a, Best-fit thermal inertia. b, Uncertainty of the thermal inertia. c, Best-fit dielectric constant. d, Uncertainty of the dielectric constant. e, Goodness-of-fit in units of $\chi_r^2$ (Equation \ref{eq:chi2}). f, Surface coverage (that is, how many times how many times a surface area was observed by ALMA). In panel a, the dashed box indicates the Bravo-Golf region discussed in Sections \ref{sec:analysis_crater} and \ref{sec:mass_deficit}. In panel c we indicate 10 areas whose surface material has median dielectric constant at least 1 standard deviation higher (squares, labelled 1--5) or lower (circles, labelled 6--10) than the median global value of Psyche. For the significance of these comparisons and their interpretation, see Section \ref{sub:metal_silicate}. The names of surface area in panel f are from \citeA{shepard2021}. 3-D animations of panels a and c are available as supporting information (videos S1 and S2, respectively). \label{fig:thermo_maps}}
\end{figure}
\begin{figure}
\centering
\includegraphics[width=\textwidth]{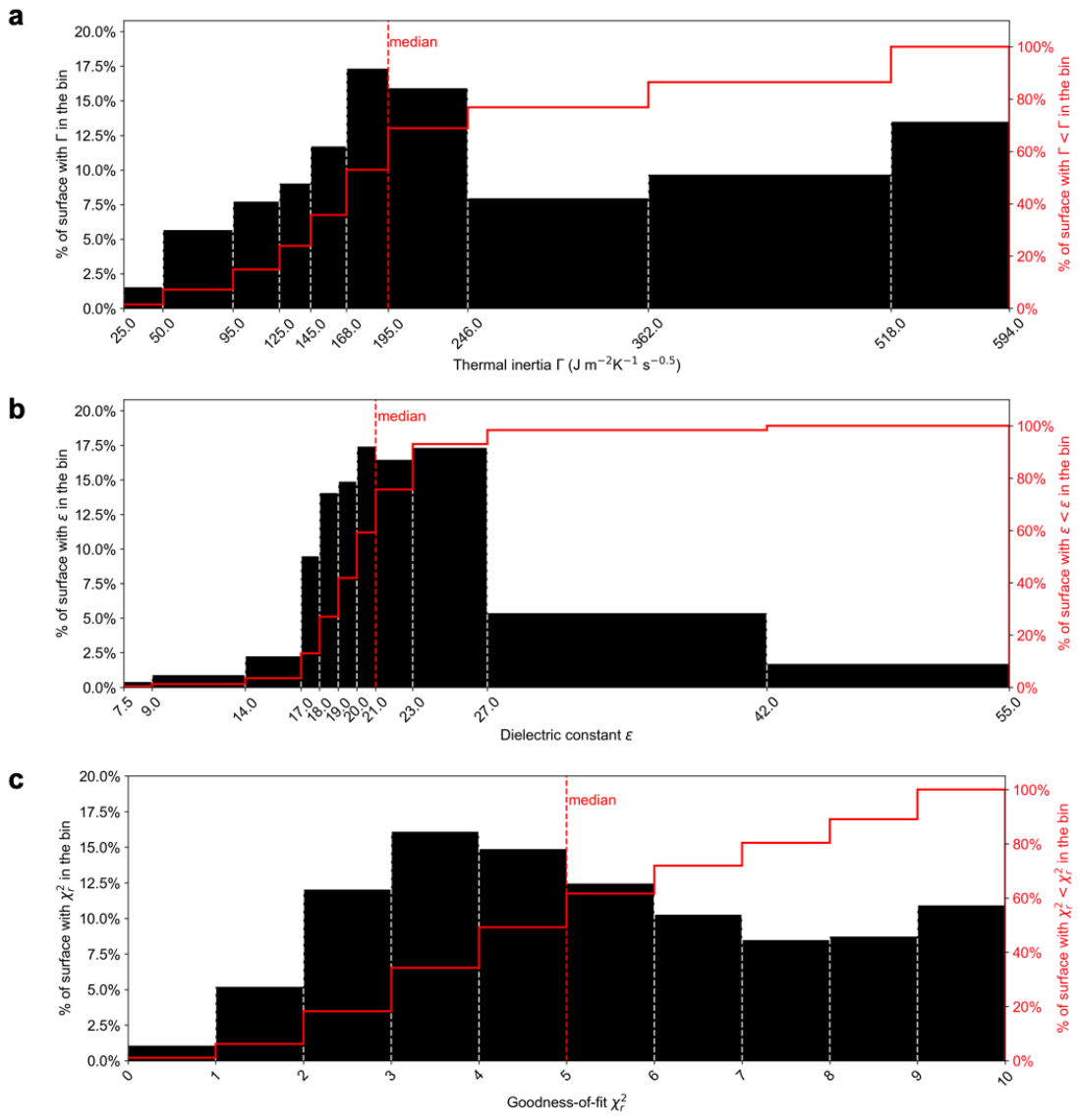}
\caption{\textbf{Most of Psyche’s surface has thermal inertia $\sim$ 150–300 J m$^{-2}$ K$^{-1}$ s$^{-0.5}$ and dielectric constant $\sim$ 15–25, consistent with the global values presented by \citeA{deKleer_2021_Psyche}.} a, Thermal inertia. b, Dielectric constant. c, Goodness-of-fit. The black histograms display areal abundance of surface area with $\Gamma$, $\epsilon$ or $\chi_r^2$ in the bin. The values are those from the maps of Figure \ref{fig:thermo_maps}, weighted according to the corresponding area on the best-fit ellipsoid by \citeA{shepard2021}. The red curves are the corresponding cumulative distributions. For panel a and b, the bin edges corresponds to the midpoints between consecutive values of $\Gamma$ and $\epsilon$ explored with the thermophysical model (Section \ref{sec:method}).  \label{fig:histograms}}
\end{figure}

Figure \ref{fig:thermo_maps} displays the maps of Psyche's surface properties derived from the ALMA data: thermal inertia $\Gamma$ (best-fit value: panel a; uncertainty: panel b); dielectric constant $\epsilon$ (best-fit value: panel c; uncertainty: panel d); the goodness-of-fit of the local solution (panel e); and the surface coverage achieved with ALMA (panel f), that is, how many times a surface area was observed by ALMA. The areal abundance of surface areas with a certain value of thermal inertia, dielectric constant and goodness-of-fit is displayed in Figure \ref{fig:histograms}, together with the corresponding cumulative distributions (red curves). Most areas have $\chi_r^2<$5--6, which we consider satisfactory for these types of observations.

We find that the surface thermal inertia of Psyche ranges between $\sim$25 tiu and $\sim$600 tiu, although most of the surface has a thermal inertia in the range 150--300 tiu (Figures \ref{fig:thermo_maps}a and \ref{fig:histograms}a). To the west of longitude 30$^\circ$W, we observe that the uncertainty of the thermal inertia measurements (Figure \ref{fig:thermo_maps}b) are comparable to the best-fit values of the thermal inertia, that is, the measurements are not well constrained, while the thermal inertia measurements are better constrained in most of the Eastern hemisphere. This is probably because the Western hemisphere was observed a fewer number of epochs (and hence times of day) than the Eastern hemisphere (Figure \ref{fig:thermo_maps}f). In areas with thermal inertia $\Gamma>$ 300 tiu, the best-fit thermal inertia tend to have higher uncertainties because the diurnal curves of the thermal emission become more and more alike as the surface thermal inertia increases.

We find that the dielectric constant of Psyche's surface material ranges between $\sim$8 and $\sim$60 (Figures \ref{fig:thermo_maps}b and \ref{fig:histograms}b), although most of the surface has dielectric constant in the range 15--25, consistent with the global solution by \citeA{deKleer_2021_Psyche}. Analogously to the thermal inertia measurements, the highest uncertainties for the dielectric constant are observed in the Western hemisphere likely because of lower surface coverage (Figure \ref{fig:thermo_maps}f). Overall, the measured dielectric constant values have much lower uncertainties than the thermal inertia values. 

\subsection{Thermal signature of a large mass-deficit region}
\label{sec:analysis_crater}
\begin{figure}
\centering
\includegraphics[width=\linewidth]{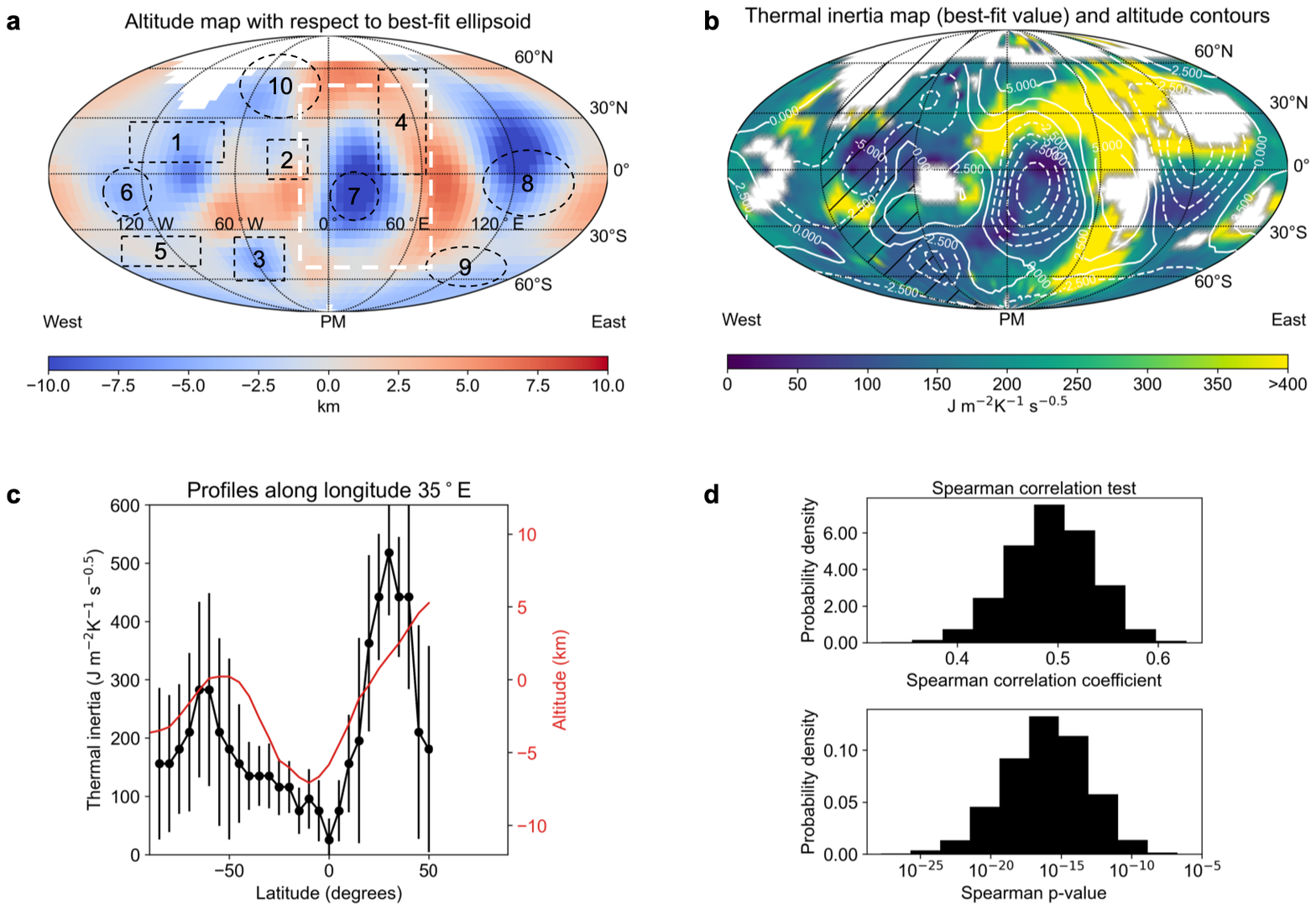}
\caption{\textbf{The lowlands in the Bravo-Golf region (longitudes 15$^\circ$W to 60$^\circ$E) have a lower thermal inertia than the surrounding highlands.} a, Altitude map computed by subtracting the best-fit ellipsoid of Psyche from its shape model by \citeA{shepard2021}. The dashed white box indicates the Bravo-Golf region analyzed in Section \ref{sec:analysis_crater} and discussed in Section \ref{sec:mass_deficit}. We also indicate the same 10 areas of Figure \ref{fig:thermo_maps}c whose surface material has median dielectric constant at least 1 standard deviation higher (squares, labelled 1--5) or lower (circles, labelled 6--10) than the median global value of Psyche (Section \ref{sub:metal_silicate}). b, Altitude contours superimposed on the map of best-fit thermal inertia of Figure \ref{fig:thermo_maps}a. c, Latitudinal profiles at longitude 35$^\circ$E of thermal inertia and altitude (black and red curves, respectively). d, Results of the two-sided Spearman test of the null hypothesis that a random distribution of thermal-inertia- and elevation-values could produce the observed correlation of panel b when uncertainties in thermal inertia are considered, for values in-between longitudes 15$^\circ$W and 60$^\circ$E and latitudes $\pm$50$^\circ$. The Spearman p-value is lower than $10^{-5}$ in all the 10,000 trials, which confirms that the thermal inertia is correlated with altitude in the region. \label{fig:crater}}
\end{figure}

%
In the Eastern hemisphere (where the values of thermal inertia are better constrained), we observe a prominent feature in the thermal inertia map extending between longitudes 15$^\circ$W and 60$^\circ$E (that is, the region in the dashed white box in Figures \ref{fig:thermo_maps}a and \ref{fig:crater}a). The feature is a low-thermal-inertia region surrounded by a ring of high-thermal inertia areas and corresponds to the Bravo and Golf regions in \citeA{shepard2021}. We refer to this region as Bravo-Golf hereafter. In Figure \ref{fig:crater}a, we plot the altitude map of Psyche computed by subtracting Psyche's best-fit ellipsoid to its shape model. Figure \ref{fig:crater}b is the same map of thermal inertia of Figure \ref{fig:thermo_maps}a with the altitude contours superimposed onto it. In Figure \ref{fig:crater}c, we plot both a thermal-inertia profile and an altitude profile (black and red curves, respectively) taken along longitude 35$^\circ$E. We observe that the two profiles closely resemble one another, and that the thermal inertia of the lowlands is statistically distinct from that of the highlands. 

To take into account the uncertainties in the value of thermal inertia, we perform a two-sided Spearman test to try the null hypothesis that a random distribution in thermal inertia and elevation in-between longitudes 15$^\circ$W and 60$^\circ$E and latitudes $\pm$50$^\circ$ could reproduce the observed feature. We perform the test 10,000 times, where at each trial we draw sample from a Gaussian distribution with mean and standard deviation equal to the best-fit $\Gamma$ and its uncertainty. We find that the correlation between altitude and thermal inertia in the region is statistically significant as the null hypothesis has a probability below 10$^{-5}$ in each trial (Figure \ref{fig:crater}d). We discuss the thermal-inertia feature in Section \ref{sec:mass_deficit}.

%

\subsection{Robustness of the results against model assumptions}
\label{sec:robustness}

We test the robustness of the results to the assumed value of Bond albedo (Section \ref{sec:method_images}). Variegations in albedo on Psyche were observed by \citeA{viikinkoski2018}, who assumed that the albedo could vary up to $\pm$ 25\% of the nominal value --- a reasonable range given that the problem of extracting albedo variations from lightcurves is poorly constrained. Using instantaneous thermal equilibrium, \citeA{deKleer_2021_Psyche} demonstrated that a variation of $\pm$ 30\% in Bond albedo is too small to explain the magnitude of the time-averaged temperature residuals in Figure \ref{fig:data_model}c. We confirm this by running additional TPM simulations for the global model of \citeA{deKleer_2021_Psyche}, but with 30\% higher/lower albedo than the nominal case. While we cannot exclude that the thermal anomalies are in part contributed by differences in temperature driven by albedo variations, we conclude that the anomalies are mainly due to heterogeneity in the thermal inertia and dielectric constant of the surface materials.

\begin{figure}
\centering
\includegraphics[width=\linewidth]{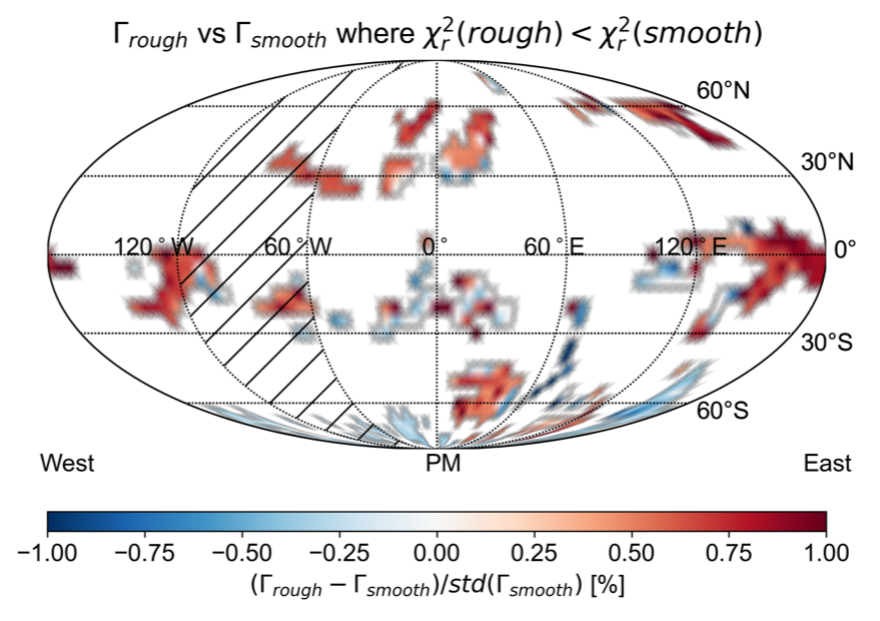}
\caption{\textbf{Tests of robustness of the results.} Relative difference between the thermal inertia $\Gamma(rough)$ obtained using the rough model and thermal inertia $\Gamma(smooth)$ obtained using the smooth model, normalized to the standard deviation of the latter. The map plots only those (few) areas where the rough solution has better goodness of fit than the smooth solution (that is, $\chi_r^2(rough)<\chi_r^2(smooth)<$ 10).}
\label{fig:rough_model} 
\end{figure}

Next, we test the robustness of the results to the assumption of smooth surface (that is, $f$ = 0$^\circ$, Section \ref{sec:method_images}). We run additional simulations in which we carve on each facet an hemispherical crater with area equal to 60\% of that of the facets (``rough model''). This corresponds to a root-mean-square slope of $f \sim$ 40$^\circ$ whose effect on emissivity is modelled as done in \citeA{deKleer_2021_Psyche}. We experiment with $\Gamma =$ (116, 283, 442) tiu, which correspond to the lower quartile in Figure \ref{fig:histograms}a, the best-fit global value from \citeA{deKleer_2021_Psyche}, and the upper quartile in Figure \ref{fig:histograms}a, respectively. Figure \ref{fig:rough_model} is a map of those areas on Psyche where the rough model has a better goodness of fit than the smooth model, that is, $\chi_r^2 (rough)<\chi_r^2 (smooth)$ (Eq. \ref{eq:chi_stat}). We find that $\chi_r^2 (rough)>$ 10 for most of the surface and that, in the few areas where $\chi_r^2 (rough)<\chi_r^2 (smooth)$, the difference between $\Gamma(rough)$ and $\Gamma (smooth)$ is within the uncertainty of $\Gamma(smooth)$. The thermal-inertia solution in the Bravo-Golf region has $\chi_r^2 (smooth)<\chi_r^2 (rough)$ in most of the lowlands and $\chi_r^2 (rough)>$ 10 in the highlands, suggesting that the thermal-inertia signature analyzed in Section \ref{sec:analysis_crater} is unlikely to be due to unmodelled roughness. We conclude that a smooth surface is a good assumption not only globally, but also locally.

\section{Discussion and interpretation of the results}
\label{sec:discussion}

In Section \ref{sub:metal_silicate}, we interpret the variations of Psyche's dielectric constant in terms of metal content of the uppermost few millimeters of regolith. In Section \ref{sec:mass_deficit}, we discuss the thermal signature of the Bravo-Golf region (that is, the region delimited by the dashed white box in Figures \ref{fig:thermo_maps}a and \ref{fig:crater}a) and propose three scenarios that may explain it.

\subsection{Relative abundance of metals and silicates over the surface}
\label{sub:metal_silicate}

The thermal inertia values of most of Psyche's surface ($\Gamma\sim$ 150--300 tiu) is consistent with a metal-rich surface as previously noted by \citeA{matter2013} and \citeA{deKleer_2021_Psyche}. Our results are also consistent with the low-emissivity case from \citeA{landsman2018}, who measured a thermal inertia of 100–200 tiu for an emissivity of 0.5 and 0.7, although their high-emissivity case finds a lower thermal inertia of 5--25 tiu. The dielectric constant of most of Psyche's surface ($\epsilon\sim$ 15--25) is also consistent with a metal-rich surface because it is higher than the typical range of Earth rocks ($\epsilon \sim$ 2--10) and closer to that of meteorites with high metal content \cite<e.g.,>{campbell1969}. 

Intriguingly, we do not observe a global correlation between thermal inertia and dielectric constant on Psyche. Interpreting this finding requires new laboratory measurements of thermal conductivity of metal-rich particulates and their dielectric constants at ALMA-relevant frequencies. Numerous laboratory measurements of the dielectric constant of terrestrial rocks and meteorites exist in the range 20 MHz–35 GHz \cite{parkhomenko1967,campbell1969,hickson2020}. By contrast, laboratory data for rocks at millimeter wavelengths ($\sim$ 100–300 GHz) are very limited \cite<e.g.,>{brouet2014}, and there are not data for powdered metals. 

While Psyche's surface appears to be predominantly metal-rich, we also observe areas whose dielectric constant ranges between $\sim 8$ to $\sim 60$. Assuming that the metal on Psyche is in the form of metallic inclusions \cite{deKleer_2021_Psyche}, $\epsilon=$ 60 corresponds to porosity lower than 25\% and a metal content $>40\%$, while the porosity and metal content are unconstrained for $\epsilon\sim$ 5. These calculations assume no reduction in emissivity due to volume scattering (see Figure 8c in \citeA{deKleer_2021_Psyche}). As such, the dielectric constant measurements (Figure \ref{fig:thermo_maps}c) suggest that the relative abundance of metals and silicates of the first millimeters of regolith varies across the surface. 

We observe that the regions labelled as n. 1, 2, 3, 4 and 5 in Figure \ref{fig:thermo_maps}c and \ref{fig:crater}c and centered at $\sim$ (15$^\circ$N, 100$^\circ$W), (10$^\circ$N, 15$^\circ$W), (45$^\circ$S,  60$^\circ$E), (30$^\circ$N, 60$^\circ$E), and (40$^\circ$S,  120$^\circ$W) have dielectric constant $\sim$ 4.9, 2.6, 2.6, 1.0, and 4.9 standard deviations higher than the median dielectric constant of Psyche, respectively. This is indicative of a locally higher metal content of the first millimeters of the surface. Intriguingly, areas n. 1, 2, and 3 are in proximity of large depressions of Psyche's shape \cite<Figure \ref{fig:crater}a and >{shepard2021} where ferrovolcanism, that is, eruption of metallic lava flows on the surface, could have preferentially occurred \cite{2019GeoRLAbrahams,2020NatJohnson}. \citeA{2019GeoRLAbrahams} showed that molten iron may erupt on the surface of a pure iron-nickel remnant. If there is a silicate layer on top of the core, the iron melt may create intrusive diapirs which can be successively uncovered through excavation by impacts \cite{2020JGRERaducan}. \citeA{2020NatJohnson} showed that, if Psyche is a differentiated object with a relatively thin silicate mantle and iron core, volatile-rich molten iron could be injected into the mantle and erupt on the surface during its cooling. They do not provide specific predictions on the eruption sites, but suggest that these are more favoured where the crust/mantle are the thinnest, which could be associated with large depressions.

The areas labelled as n. 6, 7, 8, 9 and 10 in Figure \ref{fig:thermo_maps}c and \ref{fig:crater}c and centered at $\sim$ (10$^\circ$S,  130$^\circ$W), (15$^\circ$S,  15$^\circ$E), (5$^\circ$S,  130$^\circ$E), (50$^\circ$S,  120$^\circ$E) and (50$^\circ$N,  50$^\circ$W) have dielectric constant $\sim$ 1, 1,   1.5, 1.5, and 4.4 standard deviations lower than the median dielectric constant of Psyche, respectively. This is indicative of a locally lower metal content of the first millimeters of the surface. The surface materials at these locations could be a mixture between metal- and silicate-rich materials. The presence of silicate materials on Psyche in the form of orthopyroxenes was proposed by \citeA{sanchez2017} to explain the detection of the 0.93-$\mu m$ band in the visible near-infrared spectra of Psyche. Additionally, \citeA{hardersen2005}, \citeA{2010IcarFornasier} and \citeA{takir2017} detected the presence of the 3-$\mu m$ band at longer wavelengths, which they associated with OH- or H$_2$O-bearing phases, possibly in the form of phyllosilicates \cite{takir2012}. The carbonaceous materials are likely exogenic \cite<e.g.,>{shepard2015,avdellidou2018}, consistent with experiments showing that hydrated projectiles may implant the 3-$\mu m$ band on metal-rich surfaces \cite{2019SciALibourel}. Exogeneous silicates are therefore a likely contributor to the observed areas with low dielectric constant, but we cannot conclude that they are all predominantly carbonaceous, because dielectric constant is a poor discriminator of composition between different silicate-rich materials \cite{1975EBrecher,2015IcarPalmer}. Area n. 10 has the lowest dielectric constant on Psyche ($\epsilon\sim$ 8) and could be particularly rich in silicate materials within the first millimeters of the surface. However, we caution that model artifacts could be present at that location because the region was imaged in less than 1/3 of the ALMA observations and always at high emission angles.

Assuming that Psyche's surface material is a mixture of rocks and iron sulfides/oxides \cite<Eq. 15 in>{deKleer_2021_Psyche}, the range of dielectric constant on Psyche corresponds to a local bulk density ranging from $\rho\sim$ 1500 kg/m$^3$ (i.e., 100 wt.\% iron sulfides/oxides and 67\% macroporosity) to 4500 kg/m$^3$ (i.e., 97 wt.\% iron sulfides/oxides and 0\% macroporosity). This density range encompasses the radar-derived $\rho$ = 3500 kg/m$^3$ (which we assumed for Equation \ref{eq:d_thermal}) and may also be consistent with the full density range inferred from the full range of measured radar albedos by \citeA{shepard2021}. However, we do not observe a spatial correlation between dielectric constant and radar albedo on Psyche. For example, area n. 3 \cite<India in>{shepard2021} is a region of both high dielectric constant and elevated radar albedo, but area n. 9 \cite<Hotel in>{shepard2021} is a radar-bright area with low dielectric constant. Radar-bright regions with low ALMA dielectric constant could be metal-rich areas covered by a thin mantle of silicate-rich material or be covered in a silicate-rich regolith with embedded metal particles. We cannot disentangle these two scenarios just based on dielectric constant and thermal inertia alone because our model assumes constant density with depth. In fact, the density of the uppermost millimeters of the surface (to which ALMA is sensitive to) can be different than that of the underlying first tens of centimeters, to which the radar is sensitive to. The latter is certainly true for the Moon \cite{2017JGREHayne}. Further research is therefore warranted to adjust radar densities to observations at ALMA wavelengths. This task offers many challenges, including limitations of current techniques to convert radar albedo and dielectric constant into bulk density of the regolith, the different spatial resolutions of radar versus ALMA measurements, and the lack of measurements of the dielectric properties of metal-rich materials in both solid and particulate form at ALMA wavelengths.

\subsection{Thermal inertia of the Bravo-Golf region}
\label{sec:mass_deficit}

\begin{figure}
\centering
\includegraphics[width=\linewidth]{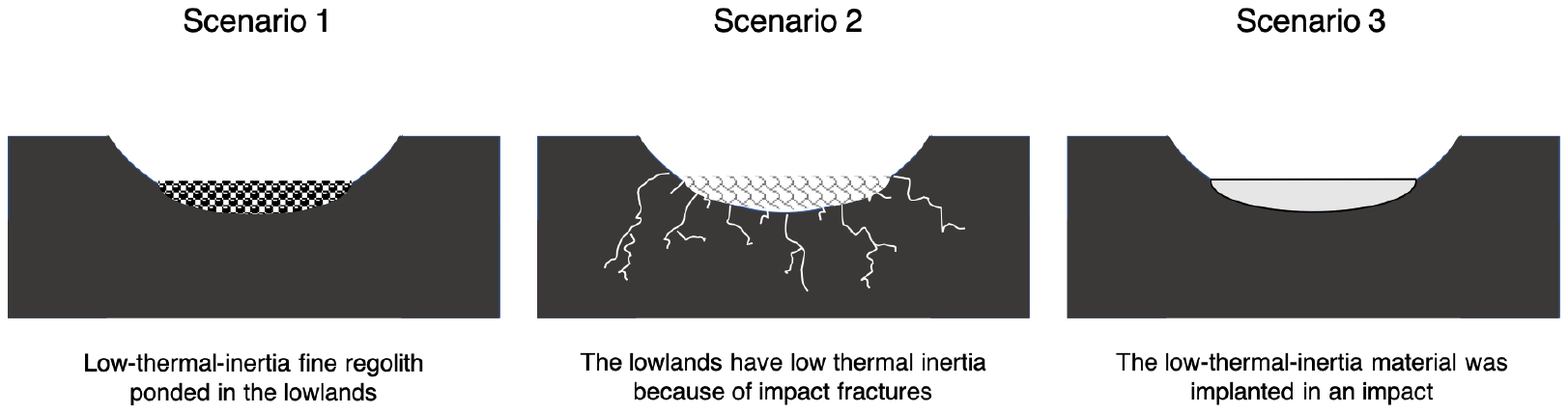}
\caption{\textbf{The three possible scenarios proposed to explain the thermal-inertia contrast between the lowlands and the highlands in the Bravo-Golf region}. The scenarios are described in Section \ref{sec:mass_deficit}. \label{fig:cartoon}}
\end{figure}

Here we propose three non--mutually-exclusive scenarios to explain why the thermal inertia of lowlands of the Bravo-Golf region is systematically lower than that of the highlands. The three scenarios are sketched in Figure \ref{fig:cartoon}.

A first hypothesis is that the lowlands are metal-rich \cite<thus radar-bright as observed by>{shepard2021}, but covered in a thin mantle of fine regolith, which lowers their thermal inertia with respect to the highlands where the regolith has coarser particle size. For a regolith particulate, thermal conductivity increases with increasing particle size for a given bulk density and specific heat capacity \cite<e.g.,>{2018IcarSakatani,cambioni2019constraining}. On airless worlds, impacts may induce accumulation of low-thermal-inertia fine regolith in places where the gravity field is the highest. The Bravo-Golf region could be a place where fine material preferentially ponds because it is closer to the center of mass of Psyche than any other point on the surface. A similar argument was presented for the bi-lobate asteroid (216) Kleopatra by \citeA{2018IcarShepard} who, based on the analysis of asteroid geopotential surface, suggested that loose material should preferentially migrate to Kleopatra's neck region, consistent with radar observations. On Psyche, we speculate that the large impacts which formed some of its depressions may have kicked-up fine regolith and/or generated seismic shaking which provoked accumulation of fines in the lowlands of the Bravo-Golf region. On asteroids greater than 100 km in diameter, individual impacts are expected to induce only localized (regional) seismic effects, but large impacts may still have widespread effects on the surface \cite{2005IcarRichardson}.

A second hypothesis is that the surface material covering the lowlands is more porous than that of the highlands. Rock's thermal inertia decreases with increasing rock porosity \cite<e.g.,>{2019NatAsGrott,cambioni2021fine}, although measurements of meteorite thermal conductivity as function of porosity are primarily based on chondrites. Rock porosity can be enhanced by impact-induced fractures. Laboratory experiments of hypervelocity impacts onto iron-nickel meteorite and ingots \cite{1984LPSCMatsui, 2019SciALibourel, 2020JGREMarchi} show that the fractures permeate the floor of the crater, while the rims acquire a rose-petal-shaped configuration and are not fractured, although results could be affected by edge effects in the samples. If the fractures extend to the entire body, this may potentially explain the low density of some M-type asteroids \cite<e.g., Psyche and Kleopatra,>{elkins-tanton2020,2021AAMarchis}. Alternatively, the surface could be covered in highly porous boulders, whose thermal inertia could mimic that of fine regolith \cite{rozitis2020asteroid,2019NatAsGrott,2020NaturOkada,cambioni2021fine}. This scenario, however, is unlikely for Psyche if its surface material has a composition analogous to that of iron meteorites \cite<porosity $<$10\%,>{macke2010survey} or mesosiderites \cite<porosity $\sim$ 12\%, similar to the $\sim$ 14\% of LL chondrites, but much lower than $\sim$ 30\% of CM chondrites,>{2003MBritt}.

A third hypothesis is that the lowlands have a higher abundance of silicate-rich materials than the highlands, consistent with having lower dielectric constant than some areas of the highlands (e.g., area 7 versus area 4 in Figure \ref{fig:thermo_maps}c). The silicate-rich materials could be the residue of a silicate impactor that may have formed the Bravo-Golf depression and whose material was mixed with the \cite<metal-rich, thus radar-bright,>{shepard2021} subsurface material by subsequent smaller impacts. If that is the case, we speculate that Psyche's surface was behaving in a ductile way at the time of impact that formed the Bravo-Golf depression, because projectile implantation efficiency is close to zero for most cooled impacts \cite{2020JGREMarchi}, assuming a target composition as that of iron meteorites. We nevertheless caution that impact contamination is unlikely to be limited to the crater floor \cite{2019SciALibourel,2020JGREMarchi} and the highlands have lower albedo than the lowlands \cite{viikinkoski2018}, in contrast with the findings by \citeA{2019SciALibourel} that silicate implantation lowers the albedo of the crater with respect to exposed iron.

\section{Conclusion}
\label{sec:conclusion}

We present a new analysis of the ALMA thermal emission data of asteroid (16) Psyche published in \citeA{deKleer_2021_Psyche} using a model which derive a value of thermal inertia and dielectric constant for each region on the surface. We find that most of Psyche's surface has thermal inertia in-between 150--300 tiu and dielectric constant 15--25.

Our results support the metal-rich nature of Psyche's surface materials, consistent with previous studies \cite{matter2013}. However, because our data spatially resolve the surface, we also observe heterogeneity in the surface properties due to the presence of terrains whose thermal inertia range between $\sim$25 tiu and $\sim$600 tiu and dielectric constant between $\sim$8 and $\sim$60. However, we do not observe a direct correlation between thermal inertia and dielectric constant on Psyche's surface. The array of detected dielectric constant values is indicative of a mineralogy of the first millimiters of the surface ranging from low-metal to high-metal content depending on porosity. Some metal-rich areas with high dielectric constant correspond to depressions where ferrovolcanism  \cite{2019GeoRLAbrahams,2020NatJohnson} could have preferentially occurred. Other regions with low dielectric constant could have higher abundance of silicate materials mixed with metals. This is consistent with spectroscopic evidence of orthopyroxenes on Psyche \cite{sanchez2017} and hydrated minerals \cite{takir2017} of likely exogeneous origin \cite{avdellidou2018}. In the thermal inertia map, we observe that the lowlands of a large depression (dubbed Bravo-Golf and located between longitudes 15$^\circ$W and 60$^\circ$E) have statistically lower thermal inertia than the surrounding highlands. We propose that this could be a signature of a thin mantle of fine regolith covering the lowlands; presence of fractures induced by impacts; and/or silicate materials implanted by impacts. All three scenarios are indicative of a collisionally evolved surface.

In conclusion, we provide evidence that Psyche is a metal-rich asteroid whose surface is heterogeneous, shows both metal and silicate materials, and appear evolved by impacts. We look forward to the data from the NASA Psyche mission to get new insight into this unusual and fascinating world.

\acknowledgments
This paper makes use of the following ALMA data: ADS/JAO.ALMA\#2018.1.01271.S. ALMA is a partnership of ESO (representing its member states), NSF (USA) and NINS (Japan), together with NRC (Canada), MOST and ASIAA (Taiwan), and KASI (Republic of Korea), in cooperation with the Republic of Chile. The Joint ALMA Observatory is operated by ESO, AUI/NRAO and NAOJ. The National Radio Astronomy Observatory is a facility of the National Science Foundation operated under cooperative agreement by Associated Universities, Inc. This research was funded in part by the Heising-Simons Foundation through grant 2019-1611. S.C. acknowledge funding through the Crosby fellowship of the Department of Earth, Atmospheric and Planetary Sciences of the Massachusetts Institute of Technology. The authors thank M. Delbo and B. Weiss for insightful discussions that improved this manuscript.\\
~\\
\noindent
\textbf{Conflict of Interest Statement}\\
\noindent
The authors declare no competing interests.\\
~\\
\noindent
\textbf{Data Availability Statement}\\
\noindent
The ALMA data are freely available at the ALMA data archive (\url{https://almascience.nrao.edu/alma-data}) under the project code: 2018.1.01271.S. The paper makes use of the ThermoPhysical Model by \citeA{delbo2015}, which is publicly available at \url{https://www.oca.eu/en/marco-delbo} in the section ``Asteroid thermal models'', and the software package PyVista by \citeA{Sullivan2019}.  All the relevant equations and parameters for the reproducibility of the results are given in this paper and in \citeA{deKleer_2021_Psyche}. Datasets for this research are available in these in-text data citation references: results in \citeA{dataset_Cambionietal_2022}; shape model in \citeA{shepard2021} (direct link in caption of their Figure 7).
\clearpage

\bibliography{Cambioni_bibliography.bib}

\end{document}


%
%


\title{Supporting Information for "The Heterogeneous Surface of Asteroid (16) Psyche"}
%
%

%
%



\authors{Saverio Cambioni\affil{1,2}, Katherine de Kleer\affil{2}, Michael Shepard\affil{3}}

\affiliation{1}{Department of Earth, Atmospheric \& Planetary Sciences, Massachusetts Institute of Technology, Cambridge, MA, USA}
\affiliation{2}{Division of Geological \& Planetary Sciences, California Institute of Technology, Pasadena, CA, USA}
\affiliation{3}{Department of Environmental, Geographical \& Geological Sciences, Bloomsburg University, Bloomsburg, PA, USA}

%
%

%

\begin{article}

%
%

\noindent\textbf{Contents of this file}
\begin{enumerate}
\item Table S1
\end{enumerate}

\noindent\textbf{Introduction}

This file contains a large table of the properties of some M-type asteroids.


\noindent\textbf{Text S1.}
%


\noindent\textbf{Data Set S1.} 


\noindent\textbf{Movie S1.} 


\noindent\textbf{Audio S1.} 
\end{article}
\clearpage